\documentclass{vgtc}                          

\ifpdf
  \pdfoutput=1\relax                   
  \usepackage{graphicx}                
  \DeclareGraphicsExtensions{.pdf,.png,.jpg,.jpeg} 
\else
  \usepackage{graphicx}                
  \DeclareGraphicsExtensions{.eps}     
\fi%

\graphicspath{{figures/}{pictures/}{images/}{./}} 

\usepackage{microtype}                 
\PassOptionsToPackage{warn}{textcomp}  
\usepackage{textcomp}                  
\usepackage{mathptmx}                  
\usepackage{times}                     
\usepackage{cite}                      
\usepackage{tabu}                      
\usepackage{booktabs}                  
\usepackage{dirtytalk}
\usepackage{hyperref}

\onlineid{1034}

\vgtccategory{PacificVis 2024 Posters}

\vgtcinsertpkg

\title{Vivifying LIME: Visual Interactive Testbed for LIME Analysis
}

\author{Jeongmin Rhee\thanks{e-mail: 202102667@hufs.ac.kr}\\ %
    \parbox{1.4in}{\scriptsize \centering Hankuk University of \\ Foreign Studies} %
\and Changhee Lee\thanks{e-mail: chlee0811@hufs.ac.kr}\\ %
     \parbox{1.4in}{\scriptsize \centering Hankuk University of \\ Foreign Studies} %
\and DongHwa Shin\thanks{e-mail: dhshin@kw.ac.kr}\\ %
    \parbox{1.4in}{\scriptsize \centering
    Kwangwoon University} %
\and Bohyoung Kim\thanks{e-mail: bkim@hufs.ac.kr, corresponding author
}\\ %
     \parbox{1.4in}{\scriptsize \centering Hankuk University of \\ Foreign Studies}
}

\abstract{
Explainable Artificial Intelligence (XAI) has gained importance in interpreting model predictions.
Among leading techniques for XAI, Local Interpretable Model-agnostic Explanations (LIME) is most frequently utilized as it notably helps people's understanding of complex models.
However, LIME's analysis is constrained to a single image at a time.
Besides, it lacks interaction mechanisms for observing the LIME's results and direct manipulations of factors affecting the results. 
To address these issues, we introduce an interactive visualization tool, LIMEVis, which improves the analysis workflow of LIME by enabling users to explore multiple LIME results simultaneously and modify them directly.
With LIMEVis, we could conveniently identify common features in images that a model seems to mainly consider for category classification. Additionally, by interactively modifying the LIME results, we could determine which segments in an image influence the model's classification.
} 

\begin{document}


\firstsection{Introduction}

\maketitle

An important role of Explainable Artificial Intelligence (XAI) is interpreting and visualizing the predictions of models.
Local Interpretable Model-agnostic Explanations (LIME)~\cite{ribeiro2016should} is one of the most widely used XAI techniques.
LIME is used for interpreting the predictions of complex models, being particularly helpful for interpreting model decisions on image and text data.
When interpreting the prediction of a model classifying an image, LIME shows which parts in the image are important to the prediction.
It takes a single image as input, breaks it into segments called \textit{superpixels}, analyzes the impact of each superpixel on the model's prediction, and outputs a single image representing highly influential superpixels.
Depending on which superpixel segmentation algorithm is used, the shape of the superpixels varies, and parameters such as \textit{positive\_only}, \textit{num\_features}, and \textit{hide\_rest} changes how the results are visually represented.

Despite of its usefulness in interpreting model's predictions, LIME has several limitations. First, it primarily focuses on explaining model's predictions for a single image at a time.
Moreover, adjusting the LIME parameters also changes the results for that single image.
This implies that the analytical space to explore for understanding the model can become vast.
Second, it only shows which superpixels had a significant impact on the prediction, not satisfying users' desire to see the impact of a specific superpixel. This limits users' ability to understand the model's prediction mechanism in more detail and identify specific causes of incorrect predictions.

To overcome these limitations, we introduce LIMEVis, an interactive visualization tool to expand the analysis approach of traditional LIME.
Our tool shows LIME results of the parameters users set for multiple images simultaneously, allowing users to analyze their model more effectively.
It also allows users to manually select superpixels and view the model's predictions to analyze the impact of those superpixels on the model.

\section{Task Summary}

\textbf{Task 1: Analyze multiple images simultaneously for model interpretation.} 
The first task is to analyze multiple images simultaneously. Using multiple images of same category and parameters set by users as input, users get multiple LIME results. The output allows users to analyze how the model predicts for that category. This gives users a comprehensive understanding of the model.


\textbf{Task 2: Allow users to manually select superpixels for model interpretation.}
The second task is to enable users to manually select superpixels and analyze their impact on the model's predictions. Users generates the desired image by viewing the original image and clicking on the superpixels they want to see. The generated image is then used as input to the model to compare the model's predictions. This allows users to analyze the specific role of a particular superpixel in the model's predictions.

\begin{figure*}
    \centering
    \includegraphics[width=1\linewidth]{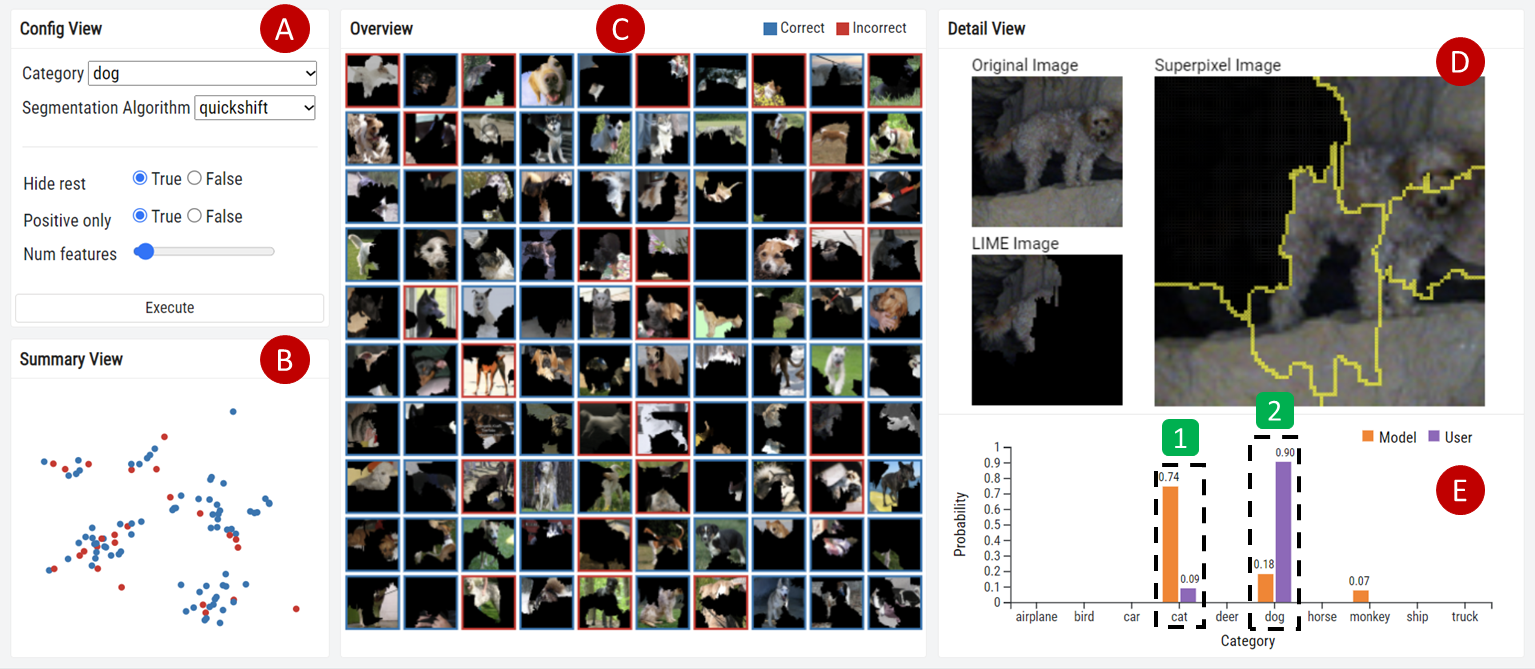}
    \vspace{-5mm}
    \caption{Visual interface of LIMEVis.
    (A) In the Config View, users can set the desired image category for which they want to interpret the model's prediction and the LIME parameters (i.e. segmentation algorithm, \textit{positive\_only}, \textit{num\_features}, and \textit{hide\_rest}).
    (B) The Summary View shows the dimensionality-reduced features of LIME results on a 2D space.
    (C) The Overview shows the original image or the LIME result images.
    (D) If users select one image they want to see in detail, three related images appear: Original Image, LIME Image, and Superpixel Image. In Superpixel Image panel, users can select superpixels by left-clicking on it.
    (E) shows the model's prediction probabilities. The purple bar represents the prediction value for the image created from user-selected superpixels. This can be compared to the orange bar that is the prediction for the original image.
    }~\label{fig:user_interface}
    \vspace{-8mm}
\end{figure*}


\section{Model}

The target model we chose to analyze through LIMEVis is an image classification model, VGG16.
The dataset we utilized is STL-10~\cite{coates2011analysis} that contains 10 categories, and each image is $96 \times 96$ pixels in size.
The training process lasted for 300 epochs with the learning rate set to 0.1. Under this setting, the test accuracy of VGG16 was 81.34\%.

\section{Visual Interface}

The system (\autoref{fig:user_interface}) consists of four main views: Config View, Overview, Summary View, and Detail View in order to provide users with the ability to analyze and interpret the model's predictions from different perspectives.

\textbf{Config View} (\autoref{fig:user_interface}-A) is where users can set the classification category and parameters for LIME analysis. There are five variables that users can set. First, users can select a classification category from the 10 categories in STL-10 dataset. 100 images belonging to the selected category will be automatically visualized. Users can also select one of three segmentation algorithms to split the superpixels. Along with this, users can set parameters such as \textit{positive\_only}, \textit{num\_features}, and \textit{hide\_rest} that affect the LIME visualization. After completing all the setting, users can click on the \say{Execute} button to run the LIME analysis.

\textbf{Overview} (\autoref{fig:user_interface}-B) displays the 100 original images belonging to the category selected in the Config View in a $10 \times 10$ format. Once the parameters have been set in the Config View and users press the \say{Execute} button, the results of the LIME analysis run with the parameters set in the Config View are displayed, again in a $10 \times 10$ format (Task 1). The borders of each image can be colored blue or red to indicate whether the VGG16 model correctly classified the image. A blue border indicates an image whose category was correctly predicted, while a red border indicates an image whose category was incorrectly predicted.

\textbf{Summary View} (\autoref{fig:user_interface}-C) shows features of 100 LIME images from the Overview as dimensionality-reduced 2D plot.
The 100 LIME Images can be seen in the Overview. Since the size of the image is too small to see the features of 100 LIME images applied at a glance in the Overview due to space limitations, we designed this view to allow users to brush images with similar features through feature extraction.
To extract features from the images, we utilized the VGG16 model pre-trained with ImageNet as a feature extractor.
Then, we use an algorithm called PacMAP~\cite{JMLR:v22:20-1061} to reduce the dimensionality of extracted image features.
The points obtained from this process are colored blue and red, just like in the Overview, to distinguish between images with correctly category predictions and those that are not. If users brush the points they want to see in the Summary View, they can see the images corresponding to the brushed points directly in the Overview.

\textbf{Detail View} is where users can perform a more in-depth analysis of one image selected in the Overview. The view consists of two parts: an interactive area at the top (\autoref{fig:user_interface}-D) and a bar chart showing the results at the bottom (\autoref{fig:user_interface}-E). At the top, users can see a total of three images. The Original Image, the LIME Image, and the Superpixel Image with the superpixels separated by a yellow border. Users can click on any part of the yellow border in the Superpixel Image panel image to choose whether to hide or show that superpixel. Clicking on a superpixel that shows an image will mask it to black, and clicking on a black-masked superpixel will show the image. Each time users select a superpixel, the bar chart at the bottom updates in real time with the VGG16 model's predictions for that image. In the bar chart, the model's prediction for the original image is colored orange, and the model's prediction for the image that changes based on user interaction is colored purple. Based on the changing purple bar, users can see how superpixels are involved in the model's predictions (Task 2).

\section{Usage Scenario}

We analyzed 100 images of the \say{dog} category using our LIMEVis system. After setting the necessary parameters for the analysis in the Config View, we could see the images with LIME applied in the Overview. To get a general idea of how the model interprets the images, we brushed the clustered points in the Summary View and viewed the corresponding images directly in the Overview. We observed that many of the images had a dog's head or tail in common. This means that the face and tail had a lot of influence on the model prediction (Task 1).

Then, we selected the images with red borders that the model made incorrect predictions and analyzed them in detail in the Detail View. In the initial prediction, the image was misclassified as \say{cat} with a probability of 0.74, and the prediction probability of \say{dog} was only 0.18 (\autoref{fig:user_interface}-1).
By analyzing the LIME image in the Detail View, it was possible to identify the superpixels that had a significant impact on the classification as \say{cat}.
Using the Superpixel Image panel, these influential superpixels were then manually disabled (Task 2), and the model's predictions were correctly classified as  \say{dog} with a probability of 0.90, and the probability of predicting \say{cat} was significantly reduced to 0.09 (\autoref{fig:user_interface}-2).

\section{Conclusion}

LIMEVis helps users better understand the factors influencing the image classification results of black box models by providing a visual interface with user interaction techniques.
Future work aims to provide modules that allow users to further revise the performance of their models by enhancing correct predictions and reducing incorrect predictions within the system pipeline.


\acknowledgments{
This work was supported by the National Research Foundation of Korea (NRF) grant funded by the Korea government (MSIT) (No. RS-2023-00251406).}

\bibliographystyle{abbrv-doi}

\bibliography{reference}
\end{document}